\newcommand*{\wn}{cm$^{-1}$}
\newcommand*{\X}{$X^1\Sigma^+_g$}
\newcommand*{\EF}{$EF^1\Sigma^+_g$}

\newcommand*{\abin}{\textit{ab initio}}

\def\apj{Astrophys.\ J. }

\def\jcp{J. Chem.\ Phys.\ }

\def\molp{Mol.\ Phys.\ }

\def\pra{Phys.\ Rev.\ A }

\def\prl{Phys.\ Rev.\ Lett.\ }

\def\jms{J. Mol.\ Spectrosc.\ }

\def\cjp{Can.\ J. Phys.\ }

\documentclass[aps,prb,twocolumn,superscriptaddress,showpacs]{revtex4}
\usepackage{graphicx}
\usepackage{amsmath}
\usepackage{comment}

\begin{document}

\title{
Test of quantum chemistry in vibrationally-hot hydrogen molecules
}

\author{M. L. Niu}
\affiliation{Department of Physics and Astronomy, and LaserLaB, VU University, De Boelelaan 1081, 1081 HV Amsterdam, The Netherlands}
\author{E. J. Salumbides}
\affiliation{Department of Physics and Astronomy, and LaserLaB, VU University, De Boelelaan 1081, 1081 HV Amsterdam, The Netherlands}
\affiliation{Department of Physics, University of San Carlos, Cebu City 6000, Philippines}
\author{W. Ubachs}
\affiliation{Department of Physics and Astronomy, and LaserLaB, VU University, De Boelelaan 1081, 1081 HV Amsterdam, The Netherlands}

\date{\today}

\begin{abstract}

\noindent
Precision measurements are performed on highly excited vibrational quantum states of molecular hydrogen.
The $v=12, J=0-3$ rovibrational levels of H$_2$ ($X^1\Sigma_g^+$), lying only $2000$ cm$^{-1}$ below the first dissociation limit, were populated by photodissociation of H$_2$S and their level energies were accurately determined by two-photon Doppler-free spectroscopy.
A comparison between the experimental results on $v=12$ level energies with the best \textit{ab initio} calculations shows good agreement, where the present experimental accuracy of $3.5 \times10^{-3}$ cm$^{-1}$ is more precise than theory, hence providing a gateway to further test theoretical advances in this benchmark quantum system.

\end{abstract}


\maketitle

Quantum chemistry started as a pioneering application of quantum mechanics, when Heitler and London explained in 1927 the existence of an attractive bonding state between the two hydrogen atoms to form the ground state of molecular hydrogen.~\cite{Heitler1927} Since then, H$_2$ has been a canonical test system, where innovations in \abin\ theory, most notably by Ko\l{}os and Wolniewicz~\cite{Kolos1968,Wolniewicz1995} and experimental measurement techniques, most notably by Herzberg and coworkers~\cite{Herzberg1959,Herzberg1970,Herzberg1972} have mutually stimulated each other as a driving force toward the highest accuracies.~\cite{Sprecher2011} Similar highly-accurate comparisons between theory and experiments are also actively pursued in molecular hydrogen ions H$_2^+$ and HD$^{+}$, where accuracies achieved in calculations are better, owing to their simpler three-body configuration~\cite{Koelemeij2007,Bressel2012,Korobov2014a}.
This simplicity on the other hand, makes the one-electron ion system to be atypical of molecules as it does not feature electron correlations, which form an essential ingredient in the quantum chemistry of molecules and condensed matter systems for which treatment of H$_2$ serves as a benchmark.

In recent years, great progress has been made in the quantum chemical calculations of the energy level structure of the H$_2$ \X\ ground electronic state.
For example, the most recent calculation of the chemical bonding energy or dissociation limit $D_0$ of the ground electronic state has an accuracy at the $10^{-3}$ \wn\ level, while nearly equally accurate binding energies were calculated for the entire manifold of rovibrational states.~\cite{Piszczatowski2009,Komasa2011}
To achieve these accuracies, the Born-Oppenheimer (BO) potential energy data points were calculated to 10$^{-9}$ \wn\ accuracy~\cite{Pachucki2010a} resulting in BO-level energies of $10^{-5}$ \wn\ accuracy. Calculations of adiabatic corrections to the BO energy~\cite{Pachucki2014} yield  accuracies of $10^{-6}$ \wn, while non-adiabatic energy corrections were determined at uncertainties at the $\sim 10^{-4}$ \wn\ level.~\cite{Pachucki2009}
At such level of precision the relativistic and radiative or quantum electrodynamic (QED) effects, amounting to several 0.1 \wn, need to be accurately accounted for, as was accomplished in the novel approach by~\citet{Komasa2011}

Direct purely vibrational transitions in the ground electronic state of H$_2$ are extremely weak due to their quadrupole nature.
They were predicted \cite{Herzberg1938} and subsequently observed by Herzberg.~\cite{Herzberg1949}
The studies of Bragg \emph{et al.},\cite{Bragg1982} using a Fourier-Transform spectrometer combined with high-pressure absorption cells exhibiting effective path lengths of 400 m, enabled the measurement of the fundamental and overtone vibrational splittings up to the (4-0) band.
This was further extended to the (5-0) band in laser-based studies~\cite{Ferguson1993} using multipass cells with effective absorption paths of 20 m.
Recent laser-based measurements have improved the experimental accuracies for the fundamental (1-0) vibrational splitting using molecular beams~\cite{Dickenson2013,Niu2014} reaching uncertainties of $1\times10^{-4}$ \wn.
Sensitive cavity-ring down laser spectroscopy was applied to the (2-0) overtone band~\cite{Campargue2012} at uncertainties of $1\times10^{-3}$ \wn, as well as on the (3-0) overtone band achieving $\sim10^{-5}-10^{-4}$ \wn\ uncertainty levels.~\cite{Cheng2012,Tan2014}

It is unlikely that the direct quadrupole excitations can be extended up to the highest vibrational quanta, since for example the (12-0) transition in H$_2$ is some six orders of magnitudes weaker than the fundamental (1-0) band.~\cite{Campargue2012}
Thus far, the H$_2$ level energies obtained from the calculational framework of Pachucki and coworkers~\cite{Piszczatowski2009,Komasa2011} are in excellent agreement with the experimental determination of the dissociation energy,~\cite{Liu2009} the fundamental vibrational splitting,~\cite{Dickenson2013} the (2-0) and (3-0) overtone band transitions,~\cite{Campargue2012,Cheng2012,Tan2014} as well the level energies of the rotational series of the lowest vibrational state.~\cite{Salumbides2011}
These highly-accurate measurements test the quantum chemical calculations of the low vibrational quantum numbers $v=0-3$.

The present experiment seeks to test the calculations of H$_2$ binding energies for vibrational levels $v=6-12$, where theoretical calculations~\cite{Komasa2011} exhibit the largest uncertainties.
We follow up on the experimental findings of Steadman and Baer,~\cite{Steadman1989} who first produced highly vibrationally excited hydrogen (H$_2$*) from the photolysis of H$_2$S. Besides the main photo-dissociation product channel yielding SH molecules, there exists an energetically allowed channel producing H$_2$* under non-equilibrium conditions, upon absorption of two UV photons in the H$_2$S molecule. They performed a single-color excitation experiment, in which the UV-pulses caused two-photon dissociation, two-photon excitation in the H$_2$ $EF-X (v',v'')$ band, and multi-photon ionization/dissociation producing H$^+$ ions. These sequential processes may involve some 6 or 7 UV photons near 290 nm depending on the excitation pathway.

\begin{figure}
\resizebox{0.42\textwidth}{!}{\includegraphics{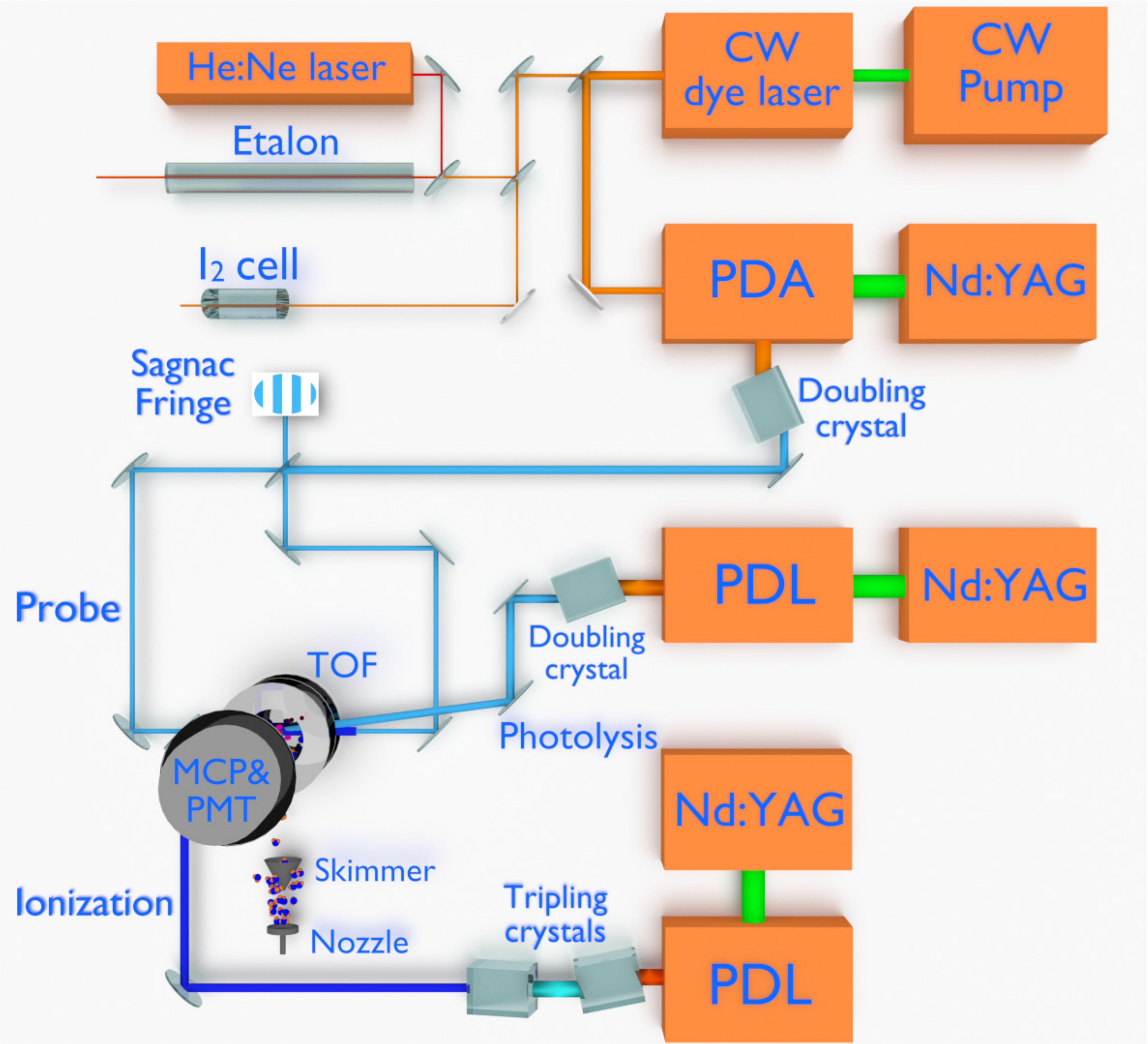}}
\caption{Schematic of the measurement setup showing the various lasers used in the photolysis  of H$_2$S (292 nm), for performing the $EF-X$ spectroscopy step  in vibrationally-hot H$_2$ (291 - 293 nm), and for inducing the ionization step (202 nm). The probe laser is split and interferometrically aligned in counter-propagating fashion to obtain Doppler-free two-photon excitation.
Resonantly produced H$^+$-ions are detected after a time-of-flight (TOF) region for resolving the ion masses.}
\label{setup}
\end{figure}

In our present study, for which a setup is displayed in Fig.~\ref{setup}, we disentangle the laser excitation processes using up to three different tunable UV-laser systems to perform a controlled study aiming at spectroscopic accuracy. In all cases, a powerful UV-laser pulse, obtained from a frequency doubled pulsed-dye laser and set at 292 nm, is used for optimal photolysis of H$_2$S  and efficient production of H$_2$*. The other laser pulses are time-delayed in steps of $\sim 10$ ns with respect to the photolysis laser to avoid temporal overlap, and therewith circumvent  broadening of the lines.
Signal is obtained by monitoring hydrogen ions, produced in the interaction region, accelerated by ion lenses and propagating over a time-of-flight (TOF) mass separation region onto a multichannel plate detection system, where the mass-resolved signal is recorded. In virtually all cases the H$^+$ signal is most prominent, although on some lines also H$_2^+$ signal is detected.

\begin{figure}
\resizebox{0.48\textwidth}{!}{\includegraphics{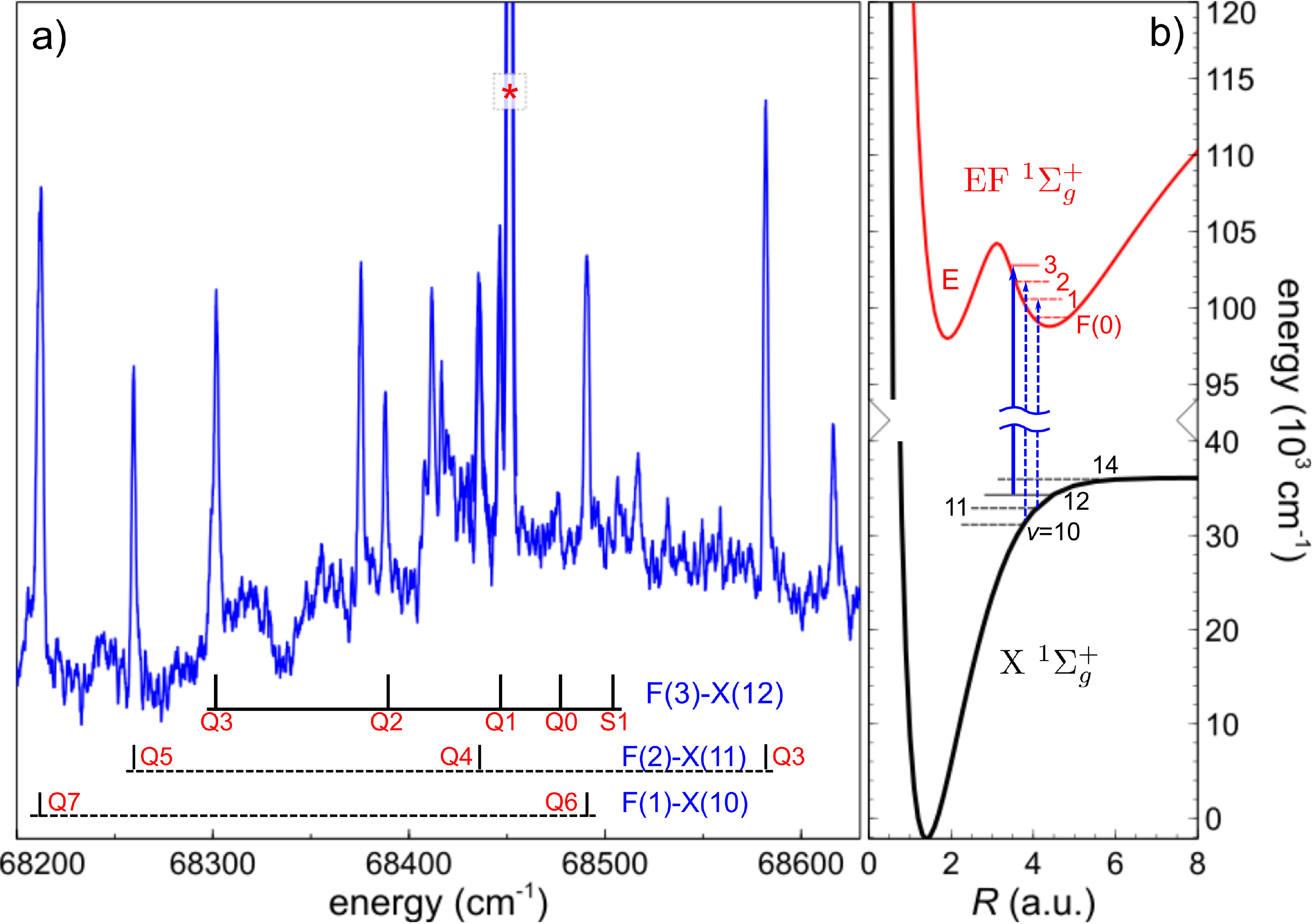}}
\caption{
a) An overview spectrum of H$_2$ showing lines assigned to $F-X (1,10)$, $F-X (2,11)$, and $F-X (3,12)$ bands. Detection is in the H$^+$ channel. The most intense line, marked by an asterisk (*) and exhibiting off-scale intensity, is unassigned.
b) Potential energy curves of \EF\ and \X\ electronic states, including two-photon transitions in the $F-X$ system probing high vibrational levels.
}
\label{overview_spectrum}
\end{figure}

As a first step of our investigation the output of a second frequency-doubled dye laser was   spatially overlapped in the interaction region with the photolysis laser. A scan over the two-photon excitation region $68\,200-68\,800$ \wn\ region is displayed in Fig.~\ref{overview_spectrum}.
The lines in this low-resolution spectrum exhibit broadening due to the laser linewidth, Doppler and ac-Stark effects, in the order of several \wn, but are narrower than in the one-color study of Steadman and Baer.~\cite{Steadman1989} At the high intensities used the ac-Stark effects also causes appreciable shifts to the transition frequencies, but the accuracy is sufficient for a tentative identification of the transitions.

In the overview spectrum the two-photon lines in the (3,12), (2,11) and (1,10) bands of the $F-X$ system were identified. Note that the numbering of vibrational levels in the $EF$ double well system can be done separately for the inner $E$ and outer $F$ well or jointly for both wells; $EF(5)$ then corresponds to $F(3)$.
These spectra and the much narrower lines measured in the Doppler-free spectra (see below) revealed many misassignments in the proposed assignments of  Steadman and Baer,~\cite{Steadman1989} although the generic finding of H$_2$* is confirmed. In particular the assignment of transitions to the $E$ inner potential well could not be confirmed, and we conclude that it is doubtful that any $E-X$ transitions were detected in their study.\cite{Steadman1989}
However, assignments of most transitions to the $F$ outer well states are confirmed presently. Based on Franck-Condon arguments the vibrationally-excited levels in the \X\ ground state can be most easily excited to the outer well $F$-levels. The potential energy diagram of Fig.~\ref{overview_spectrum}(b) illustrates that the high vibrational ground state levels $v=10-12$ have favorable Franck-Condon overlap with levels in the outer $F$ well.
In our recordings a number of lines remain unidentified, in particular the strongest resonance observed, which is marked with (*) in Fig.~\ref{overview_spectrum}.
These problems with assignments may be due to the fact that high(er) $J$-levels in the $EF$ state, in particular in the outer $F$ well, are not known to sufficient precision to establish positive identifications.

We choose the rotational series of Q transitions, \emph{i.e.} $\Delta J=0$, in the $F-X(3,12)$ band as the main focus for a precision study in H$_2$ with the aim of testing the recent quantum theoretical \emph{ab initio} calculations.~\cite{Komasa2011}
For the precision frequency measurements the beam of a UV laser, obtained from a narrowband injection-seeded traveling-wave pulsed dye amplifier (PDA), is split in two equal parts and interferometrically aligned in counter-propagating fashion, to perform two-photon Doppler-free spectroscopy. Most of the H$^+$-signal is produced by the combination of the photolysis and spectroscopy lasers. In cases where the highest spectroscopic accuracy is desired, the power of the PDA was tuned down, and a third UV laser beam at 202 nm (frequency-tripled dye laser) is used to assist ionization from the $F(3)$ state. To avoid broadening effects induced by this ionization laser, the pulse timing is again delayed by 10 ns with respect to the spectroscopy laser. In the expectation that possibly auto-ionizing resonances could help to increase signal the third laser beam was scanned over the window 202-206 nm, but no such resonances were found.

\begin{figure}
\resizebox{0.48\textwidth}{!}{\includegraphics{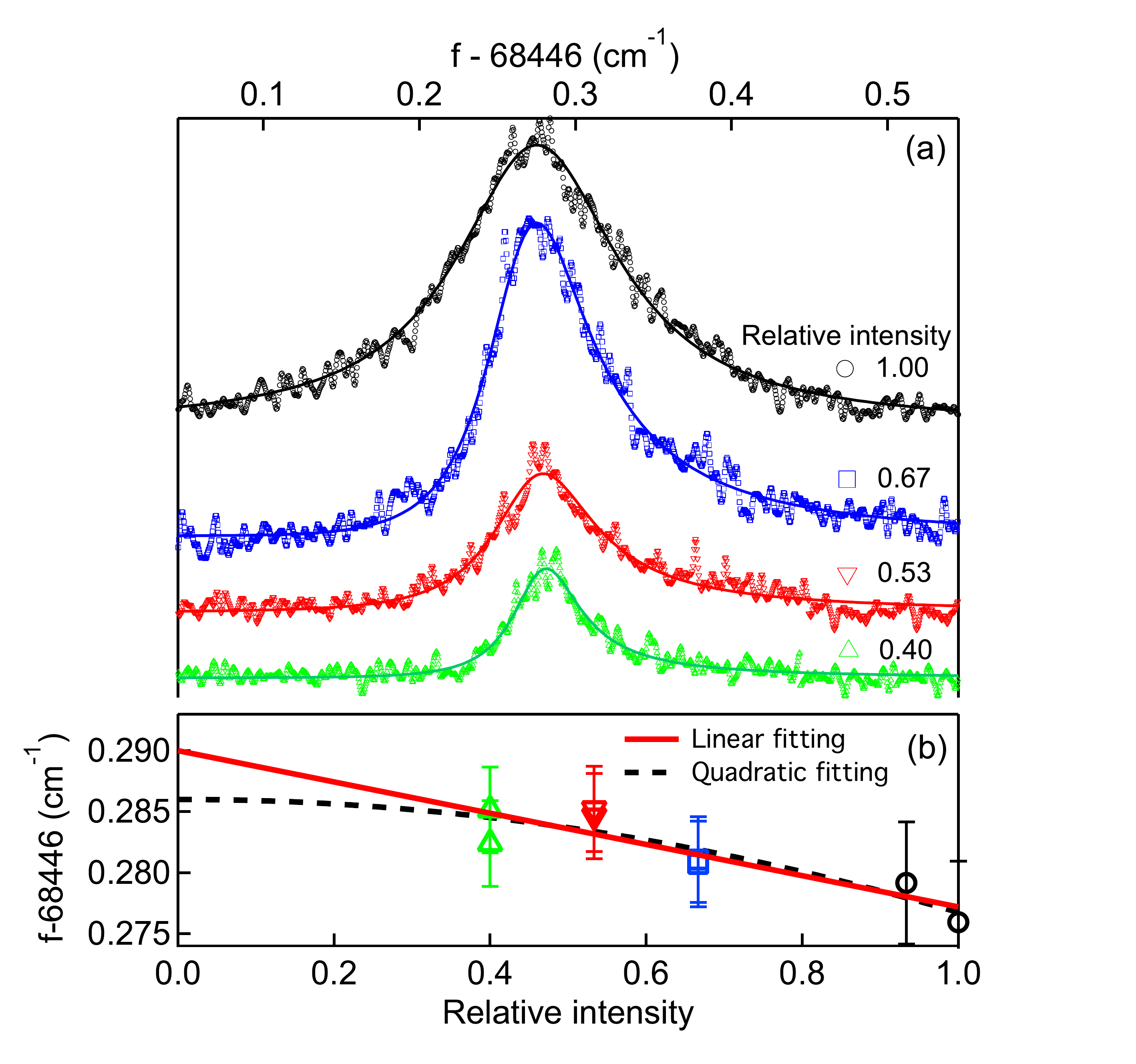}}
\caption{(a) Measurements of the Q(1) two-photon Doppler-free transition in the $F-X(3,12)$ band. (b) Plot of the peak positions at different intensities; the intercept at zero intensity (evaluated for both linear and quadratic fits), yields the (ac-Stark-free) transition frequency.}
\label{H2-PDA-spectrum-ACstark}
\end{figure}

The two-photon Q-branch lines in the $F-X(3,12)$ band were recorded under the improved high-resolution spectroscopic conditions using the narrowband PDA laser system with a bandwidth of $< 100$ MHz.
The PDA system has been described recently, including the development and characterization of a novel frequency-chirp analysis procedure to identify possible offsets between the continuous-wave seed frequency and the pulsed output of the PDA. This allows for accuracies of 0.001 \wn\ to be achieved.~\cite{Niu2015}
In Fig.~\ref{H2-PDA-spectrum-ACstark} a recording of the two-photon Q(1) transition in the $F-X(3,12)$ band is displayed, measured at different intensities of the spectroscopy laser.

For absolute frequency calibration of the cw-seed laser, hyperfine-resolved saturation spectra of I$_2$ are simultaneously recorded, using a split-off portion of the cw-seed radiation, and compared with the I$_2$ database.~\cite{Xu2000} At the same time, transmission peaks from a stabilized etalon with a free spectral range (FSR) of $150.01\,(1)$ MHz provide a relative frequency calibration, assisting to bridge the distance between H$_2$ and I$_2$ resonances.~\cite{Niu2015}
The contribution to the error budget for the H$_2$ transition frequencies from the laser scan nonlinearity, etalon calibration, and the I$_2$ calibration are estimated to be as small as 2 MHz.
The chirp-induced frequency offset between cw-seed laser and the pulsed output of the PDA system is determined and corrected for in the final frequency, yielding typical values of  $~11\,(3)$ MHz. Note that final corrections must be multiplied by a factor of 4, to account for harmonic up-conversion and two-photon excitation.

The major source of line broadening and uncertainty of the transition frequencies is the ac-Stark effect. Since the transitions are weak and the population density of H$_2^*$ low, it was necessary to use relatively high probe laser intensities, causing substantial ac-Stark broadening. With a laser bandwidth of $\sim 100$ MHz, the UV two-photon line-shapes are expected on the order of $\sim 200$ MHz, but even at the lowest laser intensities of the combined counter-propagating beams the effective line widths were $\sim 1$ GHz.
Fig.~\ref{H2-PDA-spectrum-ACstark} displays recordings of the Q(1) two-photon transition at different intensities in the interaction zone. At the lower intensities used the resonance frequency is determined from a Gaussian fit to the recorded spectrum, while at the highest intensities the resulting asymmetric lines shapes were fitted to skewed Gaussian functions of the form
$f_{SG}= (\exp{x^2/2})/(\sqrt{2\pi}) \cdot [1+{\rm erf}(x/\sqrt{2})]$.
The resulting peak positions are plotted against the relative UV intensity as shown in Fig.~\ref{H2-PDA-spectrum-ACstark}(b).
To account for the increased widths and skew, the peak position uncertainty in the high intensity scans are larger.
In view of the long extrapolation to zero intensity fits were made for a linear and a quadratic fit. Results from these approaches were averaged and the uncertainties were conservatively estimated to cover the values when differing. In Table~\ref{Table-H2-frequencies} the thus determined two-photon transition frequencies for the Q(0)-Q(3) lines in the $F-X (3,12)$ band are listed with accuracies of typically 100 MHz or 0.0035 \wn\ deriving fully from this ac-Stark analysis.

\begin{table}
\begin{center}
\begin{small}
\caption{Measured two-photon transition frequencies in the $F-X (3,12)$ band, equivalent to the $EF-X (5,12)$ band.
}
\label{Table-H2-frequencies}
\begin{tabular}{cr@{.}l@{\hspace{8pt}}r@{.}l}
\colrule
Line & \multicolumn{2}{c}{Frequency}	\\
\colrule
Q(0)	&	68\,476 & 0459\,(35)		\\
Q(1)	&	68\,446 & 2834\,(35)	    \\
Q(2)	&	68\,387 & 6623\,(35)		\\
Q(3)	&	68\,302 & 0250\,(35)		\\
S(1)    &   68\,505 & 518\,(7)           \\
\colrule
\end{tabular}
\end{small}
\end{center}
\end{table}

In order to make a comparison with the quantum chemical \abin\ calculations these values for the accurate transition frequencies must be translated into values of level energies in H$_2$ \X$(v=12, J)$.
This can be accomplished by subtracting transition energies from the accurate level energies of the upper $F(3)$ states from Bailly \emph{et al.},~\cite{Bailly2009}  which were determined via combined laser-based and Fourier-transform emission experiments yielding values accurate to within 0.001 \wn\ for $J \leq 3$. So this transformation does not add to the uncertainty of H$_2$* level energies, which are listed in Table~\ref{Table-H2-lines-results} as experimental values ($E_\mathrm{exp}$) for $v=12$.
This procedure of combining differences between $F(3)$ level energies~\cite{Bailly2009} and $X(12)$ level energies~\cite{Komasa2011}
provides a final and unambiguous assignment of the observed lines.
The measurement of a weaker transition S(1) at in the same $F-X(3,12)$ band ($\Delta J = 2$ for S transitions) allows for an additional consistency check of the line identification.
The combination difference between S(1) and Q(3) transitions delivers a level splitting between $J=1-3$ in $X(12)$ amounting to $203.494$ \wn, in agreement, at $0.008\,(8)$ \wn, from the \abin\ values.~\cite{Komasa2011} In addition, the combination difference between S(1) and Q(1) lines is in reasonable agreement with the splitting between $J=1-3$ in $F(3)$ within $0.010\,(8)$ \wn. 

\begin{table}
\begin{center}
\begin{small}
\caption{Level energies ($E_\mathrm{exp}$) of ro-vibrational states  \X\,$(v=12, J=0-3)$, experimentally determined from the measured $F-X(3,12)$ two-photon Q-transitions and the accurate $F(3)$ levels from Bailly \emph{et al.}\cite{Bailly2009}
Also indicated are the theory values ($E_\mathrm{the}$) of Komasa \emph{et al.},\cite{Komasa2011} and the difference $\Delta E_\mathrm{exp-the}$ between the experimental and theoretical values. The uncertainties in the differences represent the combined uncertainties from experiment and theory taken in quadrature.}
\label{Table-H2-lines-results}
\begin{tabular}{cr@{.}l@{\hspace{8pt}}r@{.}l@{\hspace{10pt}}r@{.}l}
\colrule
$J''$ & \multicolumn{2}{c}{$E_\mathrm{exp}$}&	\multicolumn{2}{c}{$E_\mathrm{the}$}& \multicolumn{2}{c}{$\Delta E_\mathrm{exp-the}$} \\
\colrule
0	&	34\,302	&	1823\,(35)	&	34\,302	&	174\,1\,(47)	&	0&008\,(6)	\\
1	&	34\,343	&	8531\,(35)	&	34\,343	&	848\,3\,(46)	&	0&005\,(6)	\\
2	&	34\,426	&	2216\,(35)	&	34\,426	&	217\,9\,(46)	&	0&004\,(6)	\\
3	&	34\,547	&	3362\,(35)	&	34\,547	&	333\,2\,(45)	&	0&003\,(6)	\\
\colrule
\end{tabular}
\end{small}
\end{center}
\end{table}

In Table~\ref{Table-H2-lines-results} a final comparison is made between experimentally determined level energies ($E_\mathrm{exp}$) and results from the quantum chemical calculations ($E_\mathrm{the}$) by~\citet{Komasa2011} For three of the four levels determined the difference between the experimental and theoretical values, $\Delta E_\mathrm{exp-the}$, is smaller than their combined uncertainties, while the $J''=0$ level deviated by $1.3\sigma$.
This allows for the main conclusion of the present study that good agreement is established between experiment and the quantum chemical calculations for $v=12$ levels in the electronic ground state in H$_2$. This is the highest vibrational level subjected to a precision test below 0.01 \wn\ for the H$_2$ molecule.

\begin{figure}
\resizebox{0.42\textwidth}{!}{\includegraphics{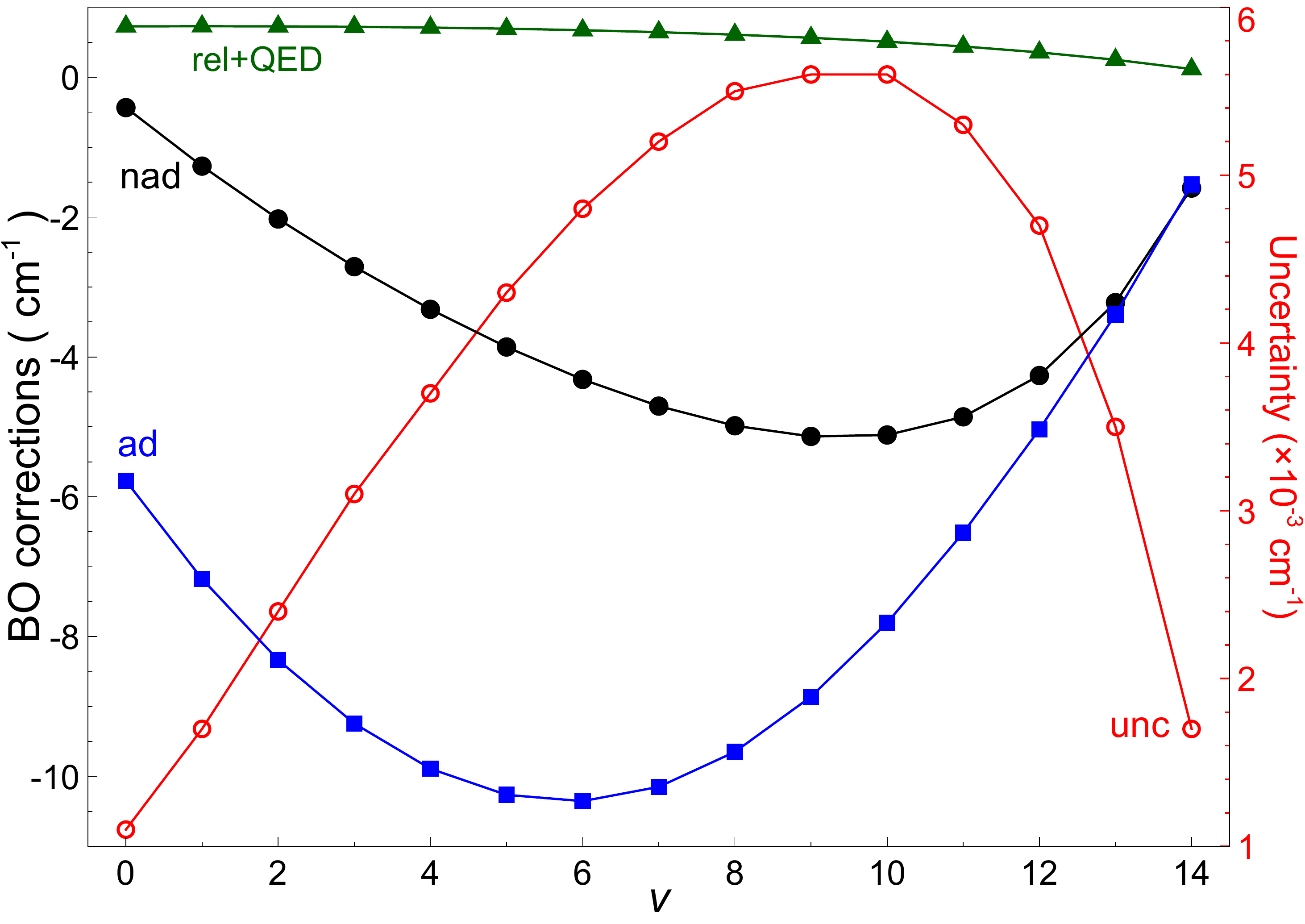}}
\caption{Plot of corrections to the \abin\ Born-Oppenheimer energies of the H$_2$ \X\ ground electronic state~\cite{Komasa2011} as a function of vibrational quantum number $v$ (in all cases $J=0$). Adiabatic (ad), nonadiabatic (nad), and  relativistic and radiative (rel+QED) corrections are plotted with reference to the scale on the left.
The total uncertainty of the calculations (unc) is plotted with reference to the scale on the right.}
\label{energy_contrib}
\end{figure}

In Fig.~\ref{energy_contrib} the contributing energy corrections to the accurate BO-energies~\cite{Pachucki2010a} are plotted as a function of $v$, as well as the final uncertainties in the theoretical binding energies.\cite{Komasa2011} It shows that the calculations  are least accurate for the range $v=6-12$. The present test on $v=12$ covers this weakest spot in the calculations.
Specifically the non-adiabatic contributions to the relativistic and QED energies (the so-called recoil corrections), as well as higher order non-adiabatic, $\alpha^4$-QED contributions, and the leading term in $\alpha^5$-QED, are mainly held responsible for possible deviations with theory.~\cite{Komasa2011} The presently determined experimental binding energies, being slightly more accurate than theory, may serve to stimulate calculations of such terms.

As an outlook from the experimental perspective, efforts are underway to determine level energies for states up to the highest possible ($v=14$) vibrational quantum states in H$_2$.
At least an order of magnitude improvement is foreseen in the experimental determination of the $v=10-14, J$ level energies, which should be achievable with an improved detection sensitivity, whereby the limiting factor of uncertainty, the ac-Stark broadening, can be suppressed.

\smallskip

\emph{Note added during review.}
After submission of this manuscript a paper reporting recalculation of the non-adiabatic interaction was published~\cite{Pachucki2015} improving their accuracy to the $10^{-7}$ \wn\ level. It was not specified in how far the newly calculated non-adiabatic wave functions affect the uncertainties of the QED and relativistic corrections.

\smallskip

This work was supported by the Netherlands Foundation for Fundamental Research of Matter (FOM) and by the Dutch Astrochemistry Network (NWO).
We thank F. Ramirez (University of San Carlos, Cebu, Philippines) for assistance in the initial investigations.

\bibliographystyle{apsrev4-1-nourl}
\bibliography{H2-references}

\begin{thebibliography}{31}%
\makeatletter
\providecommand \@ifxundefined [1]{%
 \@ifx{#1\undefined}
}%
\providecommand \@ifnum [1]{%
 \ifnum #1\expandafter \@firstoftwo
 \else \expandafter \@secondoftwo
 \fi
}%
\providecommand \@ifx [1]{%
 \ifx #1\expandafter \@firstoftwo
 \else \expandafter \@secondoftwo
 \fi
}%
\providecommand \natexlab [1]{#1}%
\providecommand \enquote  [1]{``#1''}%
\providecommand \bibnamefont  [1]{#1}%
\providecommand \bibfnamefont [1]{#1}%
\providecommand \citenamefont [1]{#1}%
\providecommand \href@noop [0]{\@secondoftwo}%
\providecommand \href [0]{\begingroup \@sanitize@url \@href}%
\providecommand \@href[1]{\@@startlink{#1}\@@href}%
\providecommand \@@href[1]{\endgroup#1\@@endlink}%
\providecommand \@sanitize@url [0]{\catcode `\\12\catcode `\$12\catcode
  `\&12\catcode `\#12\catcode `\^12\catcode `\_12\catcode `\%12\relax}%
\providecommand \@@startlink[1]{}%
\providecommand \@@endlink[0]{}%
\providecommand \url  [0]{\begingroup\@sanitize@url \@url }%
\providecommand \@url [1]{\endgroup\@href {#1}{\urlprefix }}%
\providecommand \urlprefix  [0]{URL }%
\providecommand \Eprint [0]{\href }%
\@ifxundefined \urlstyle {%
  \providecommand \doi  [0]{\begingroup \@sanitize@url \@doi}%
  \providecommand \@doi [1]{\endgroup \@@startlink {\doibase
  #1}doi:\discretionary {}{}{}#1\@@endlink }%
}{%
  \providecommand \doi  [0]{doi:\discretionary{}{}{}\begingroup
  \urlstyle{rm}\Url }%
}%
\providecommand \doibase [0]{http://dx.doi.org/}%
\providecommand \Doi [0]{\begingroup \@sanitize@url \@Doi }%
\providecommand \@Doi  [1]{\endgroup\@@startlink{\doibase#1}\@@Doi}%
\providecommand \@@Doi [1]{#1\@@endlink}%
\providecommand \selectlanguage [0]{\@gobble}%
\providecommand \bibinfo  [0]{\@secondoftwo}%
\providecommand \bibfield  [0]{\@secondoftwo}%
\providecommand \translation [1]{[#1]}%
\providecommand \BibitemOpen [0]{}%
\providecommand \bibitemStop [0]{}%
\providecommand \bibitemNoStop [0]{.\EOS\space}%
\providecommand \EOS [0]{\spacefactor3000\relax}%
\providecommand \BibitemShut  [1]{\csname bibitem#1\endcsname}%
\bibitem [{\citenamefont {Heitler}\ and\ \citenamefont
  {London}(1927)}]{Heitler1927}%
  \BibitemOpen
  \bibfield  {author} {\bibinfo {author} {\bibfnamefont {W.}~\bibnamefont
  {Heitler}}\ and\ \bibinfo {author} {\bibfnamefont {F.}~\bibnamefont
  {London}},\ }\href@noop {} {\bibfield  {journal} {\bibinfo  {journal} {Zeit.
  f. Phys.},\ }\textbf {\bibinfo {volume} {44}},\ \bibinfo {pages} {455}
  (\bibinfo {year} {1927})}\BibitemShut {NoStop}%
\bibitem [{\citenamefont {Ko\l{}os}\ and\ \citenamefont
  {Wolniewicz}(1968)}]{Kolos1968}%
  \BibitemOpen
  \bibfield  {author} {\bibinfo {author} {\bibfnamefont {W.}~\bibnamefont
  {Ko\l{}os}}\ and\ \bibinfo {author} {\bibfnamefont {L.}~\bibnamefont
  {Wolniewicz}},\ }\href@noop {} {\bibfield  {journal} {\bibinfo  {journal} {J.
  Chem. Phys.},\ }\textbf {\bibinfo {volume} {49}},\ \bibinfo {pages} {404}
  (\bibinfo {year} {1968})}\BibitemShut {NoStop}%
\bibitem [{\citenamefont {Wolniewicz}(1995)}]{Wolniewicz1995}%
  \BibitemOpen
  \bibfield  {author} {\bibinfo {author} {\bibfnamefont {L.}~\bibnamefont
  {Wolniewicz}},\ }\href@noop {} {\bibfield  {journal} {\bibinfo  {journal}
  {\jcp},\ }\textbf {\bibinfo {volume} {103}},\ \bibinfo {pages} {1792}
  (\bibinfo {year} {1995})}\BibitemShut {NoStop}%
\bibitem [{\citenamefont {Herzberg}\ and\ \citenamefont
  {Howe}(1959)}]{Herzberg1959}%
  \BibitemOpen
  \bibfield  {author} {\bibinfo {author} {\bibfnamefont {G.}~\bibnamefont
  {Herzberg}}\ and\ \bibinfo {author} {\bibfnamefont {L.~L.}\ \bibnamefont
  {Howe}},\ }\href@noop {} {\bibfield  {journal} {\bibinfo  {journal} {\cjp},\
  }\textbf {\bibinfo {volume} {37}},\ \bibinfo {pages} {636} (\bibinfo {year}
  {1959})}\BibitemShut {NoStop}%
\bibitem [{\citenamefont {Herzberg}(1970)}]{Herzberg1970}%
  \BibitemOpen
  \bibfield  {author} {\bibinfo {author} {\bibfnamefont {G.}~\bibnamefont
  {Herzberg}},\ }\href@noop {} {\bibfield  {journal} {\bibinfo  {journal} {J.
  Mol. Spectrosc.},\ }\textbf {\bibinfo {volume} {33}},\ \bibinfo {pages} {147}
  (\bibinfo {year} {1970})}\BibitemShut {NoStop}%
\bibitem [{\citenamefont {Herzberg}\ and\ \citenamefont
  {Jungen}(1972)}]{Herzberg1972}%
  \BibitemOpen
  \bibfield  {author} {\bibinfo {author} {\bibfnamefont {G.}~\bibnamefont
  {Herzberg}}\ and\ \bibinfo {author} {\bibfnamefont {C.}~\bibnamefont
  {Jungen}},\ }\href@noop {} {\bibfield  {journal} {\bibinfo  {journal} {J.
  Mol. Spectrosc.},\ }\textbf {\bibinfo {volume} {41}},\ \bibinfo {pages} {425}
  (\bibinfo {year} {1972})}\BibitemShut {NoStop}%
\bibitem [{\citenamefont {Sprecher}\ \emph {et~al.}(2011)\citenamefont
  {Sprecher}, \citenamefont {Jungen}, \citenamefont {Ubachs},\ and\
  \citenamefont {Merkt}}]{Sprecher2011}%
  \BibitemOpen
  \bibfield  {author} {\bibinfo {author} {\bibfnamefont {D.}~\bibnamefont
  {Sprecher}}, \bibinfo {author} {\bibfnamefont {C.}~\bibnamefont {Jungen}},
  \bibinfo {author} {\bibfnamefont {W.}~\bibnamefont {Ubachs}}, \ and\ \bibinfo
  {author} {\bibfnamefont {F.}~\bibnamefont {Merkt}},\ }\href@noop {}
  {\bibfield  {journal} {\bibinfo  {journal} {Faraday Discuss.},\ }\textbf
  {\bibinfo {volume} {150}},\ \bibinfo {pages} {51} (\bibinfo {year}
  {2011})}\BibitemShut {NoStop}%
\bibitem [{\citenamefont {Koelemeij}\ \emph {et~al.}(2007)\citenamefont
  {Koelemeij}, \citenamefont {Roth}, \citenamefont {Wicht}, \citenamefont
  {Ernsting},\ and\ \citenamefont {Schiller}}]{Koelemeij2007}%
  \BibitemOpen
  \bibfield  {author} {\bibinfo {author} {\bibfnamefont {J.~C.~J.}\
  \bibnamefont {Koelemeij}}, \bibinfo {author} {\bibfnamefont {B.}~\bibnamefont
  {Roth}}, \bibinfo {author} {\bibfnamefont {A.}~\bibnamefont {Wicht}},
  \bibinfo {author} {\bibfnamefont {I.}~\bibnamefont {Ernsting}}, \ and\
  \bibinfo {author} {\bibfnamefont {S.}~\bibnamefont {Schiller}},\ }\href@noop
  {} {\bibfield  {journal} {\bibinfo  {journal} {\prl},\ }\textbf {\bibinfo
  {volume} {98}},\ \bibinfo {pages} {173002} (\bibinfo {year}
  {2007})}\BibitemShut {NoStop}%
\bibitem [{\citenamefont {Bressel}\ \emph {et~al.}(2012)\citenamefont
  {Bressel}, \citenamefont {Borodin}, \citenamefont {Shen}, \citenamefont
  {Hansen}, \citenamefont {Ernsting},\ and\ \citenamefont
  {Schiller}}]{Bressel2012}%
  \BibitemOpen
  \bibfield  {author} {\bibinfo {author} {\bibfnamefont {U.}~\bibnamefont
  {Bressel}}, \bibinfo {author} {\bibfnamefont {A.}~\bibnamefont {Borodin}},
  \bibinfo {author} {\bibfnamefont {J.}~\bibnamefont {Shen}}, \bibinfo {author}
  {\bibfnamefont {M.}~\bibnamefont {Hansen}}, \bibinfo {author} {\bibfnamefont
  {I.}~\bibnamefont {Ernsting}}, \ and\ \bibinfo {author} {\bibfnamefont
  {S.}~\bibnamefont {Schiller}},\ }\href@noop {} {\bibfield  {journal}
  {\bibinfo  {journal} {\prl},\ }\textbf {\bibinfo {volume} {108}},\ \bibinfo
  {pages} {183003} (\bibinfo {year} {2012})}\BibitemShut {NoStop}%
\bibitem [{\citenamefont {Korobov}\ \emph {et~al.}(2014)\citenamefont
  {Korobov}, \citenamefont {Hilico},\ and\ \citenamefont
  {Karr}}]{Korobov2014a}%
  \BibitemOpen
  \bibfield  {author} {\bibinfo {author} {\bibfnamefont {V.~I.}\ \bibnamefont
  {Korobov}}, \bibinfo {author} {\bibfnamefont {L.}~\bibnamefont {Hilico}}, \
  and\ \bibinfo {author} {\bibfnamefont {J.-P.}\ \bibnamefont {Karr}},\ }\Doi
  {10.1103/PhysRevA.89.032511} {\bibfield  {journal} {\bibinfo  {journal}
  {\pra},\ }\textbf {\bibinfo {volume} {89}},\ \bibinfo {pages} {032511}
  (\bibinfo {year} {2014})}\BibitemShut {NoStop}%
\bibitem [{\citenamefont {Piszczatowski}\ \emph {et~al.}(2009)\citenamefont
  {Piszczatowski}, \citenamefont {\L{}ach}, \citenamefont {Przybytek},
  \citenamefont {Komasa}, \citenamefont {Pachucki},\ and\ \citenamefont
  {Jeziorski}}]{Piszczatowski2009}%
  \BibitemOpen
  \bibfield  {author} {\bibinfo {author} {\bibfnamefont {K.}~\bibnamefont
  {Piszczatowski}}, \bibinfo {author} {\bibfnamefont {G.}~\bibnamefont
  {\L{}ach}}, \bibinfo {author} {\bibfnamefont {M.}~\bibnamefont {Przybytek}},
  \bibinfo {author} {\bibfnamefont {J.}~\bibnamefont {Komasa}}, \bibinfo
  {author} {\bibfnamefont {K.}~\bibnamefont {Pachucki}}, \ and\ \bibinfo
  {author} {\bibfnamefont {B.}~\bibnamefont {Jeziorski}},\ }\href@noop {}
  {\bibfield  {journal} {\bibinfo  {journal} {J. Chem. Theory Comput.},\
  }\textbf {\bibinfo {volume} {5}},\ \bibinfo {pages} {3039} (\bibinfo {year}
  {2009})}\BibitemShut {NoStop}%
\bibitem [{\citenamefont {Komasa}\ \emph {et~al.}(2011)\citenamefont {Komasa},
  \citenamefont {Piszczatowski}, \citenamefont {\L{}ach}, \citenamefont
  {Przybytek}, \citenamefont {Jeziorski},\ and\ \citenamefont
  {Pachucki}}]{Komasa2011}%
  \BibitemOpen
  \bibfield  {author} {\bibinfo {author} {\bibfnamefont {J.}~\bibnamefont
  {Komasa}}, \bibinfo {author} {\bibfnamefont {K.}~\bibnamefont
  {Piszczatowski}}, \bibinfo {author} {\bibfnamefont {G.}~\bibnamefont
  {\L{}ach}}, \bibinfo {author} {\bibfnamefont {M.}~\bibnamefont {Przybytek}},
  \bibinfo {author} {\bibfnamefont {B.}~\bibnamefont {Jeziorski}}, \ and\
  \bibinfo {author} {\bibfnamefont {K.}~\bibnamefont {Pachucki}},\ }\href@noop
  {} {\bibfield  {journal} {\bibinfo  {journal} {J. Chem. Theory Comput.},\
  }\textbf {\bibinfo {volume} {7}},\ \bibinfo {pages} {3105} (\bibinfo {year}
  {2011})}\BibitemShut {NoStop}%
\bibitem [{\citenamefont {Pachucki}(2010)}]{Pachucki2010a}%
  \BibitemOpen
  \bibfield  {author} {\bibinfo {author} {\bibfnamefont {K.}~\bibnamefont
  {Pachucki}},\ }\href@noop {} {\bibfield  {journal} {\bibinfo  {journal}
  {Phys. Rev. A},\ }\textbf {\bibinfo {volume} {82}},\ \bibinfo {pages}
  {032509} (\bibinfo {year} {2010})}\BibitemShut {NoStop}%
\bibitem [{\citenamefont {Pachucki}\ and\ \citenamefont
  {Komasa}(2014)}]{Pachucki2014}%
  \BibitemOpen
  \bibfield  {author} {\bibinfo {author} {\bibfnamefont {K.}~\bibnamefont
  {Pachucki}}\ and\ \bibinfo {author} {\bibfnamefont {J.}~\bibnamefont
  {Komasa}},\ }\href@noop {} {\bibfield  {journal} {\bibinfo  {journal} {J.
  Chem. Phys.},\ }\textbf {\bibinfo {volume} {141}},\ \bibinfo {pages} {224103}
  (\bibinfo {year} {2014})}\BibitemShut {NoStop}%
\bibitem [{\citenamefont {Pachucki}\ and\ \citenamefont
  {Komasa}(2009)}]{Pachucki2009}%
  \BibitemOpen
  \bibfield  {author} {\bibinfo {author} {\bibfnamefont {K.}~\bibnamefont
  {Pachucki}}\ and\ \bibinfo {author} {\bibfnamefont {J.}~\bibnamefont
  {Komasa}},\ }\href@noop {} {\bibfield  {journal} {\bibinfo  {journal} {J.
  Chem. Phys.},\ }\textbf {\bibinfo {volume} {130}},\ \bibinfo {pages} {164113}
  (\bibinfo {year} {2009})}\BibitemShut {NoStop}%
\bibitem [{\citenamefont {{Herzberg}}(1938)}]{Herzberg1938}%
  \BibitemOpen
  \bibfield  {author} {\bibinfo {author} {\bibfnamefont {G.}~\bibnamefont
  {{Herzberg}}},\ }\href@noop {} {\bibfield  {journal} {\bibinfo  {journal}
  {\apj},\ }\textbf {\bibinfo {volume} {87}},\ \bibinfo {pages} {428} (\bibinfo
  {year} {1938})}\BibitemShut {NoStop}%
\bibitem [{\citenamefont {Herzberg}(1949)}]{Herzberg1949}%
  \BibitemOpen
  \bibfield  {author} {\bibinfo {author} {\bibfnamefont {G.}~\bibnamefont
  {Herzberg}},\ }\href@noop {} {\bibfield  {journal} {\bibinfo  {journal}
  {Nature},\ }\textbf {\bibinfo {volume} {163}},\ \bibinfo {pages} {170}
  (\bibinfo {year} {1949})}\BibitemShut {NoStop}%
\bibitem [{\citenamefont {Bragg}\ \emph {et~al.}(1982)\citenamefont {Bragg},
  \citenamefont {Smith},\ and\ \citenamefont {Brault}}]{Bragg1982}%
  \BibitemOpen
  \bibfield  {author} {\bibinfo {author} {\bibfnamefont {S.~L.}\ \bibnamefont
  {Bragg}}, \bibinfo {author} {\bibfnamefont {W.~H.}\ \bibnamefont {Smith}}, \
  and\ \bibinfo {author} {\bibfnamefont {J.~W.}\ \bibnamefont {Brault}},\ }\Doi
  {10.1086/160568} {\bibfield  {journal} {\bibinfo  {journal} {\apj},\ }\textbf
  {\bibinfo {volume} {263}},\ \bibinfo {pages} {999} (\bibinfo {year}
  {1982})}\BibitemShut {NoStop}%
\bibitem [{\citenamefont {Ferguson}\ \emph {et~al.}(1993)\citenamefont
  {Ferguson}, \citenamefont {Rao}, \citenamefont {Mickelson},\ and\
  \citenamefont {Larson}}]{Ferguson1993}%
  \BibitemOpen
  \bibfield  {author} {\bibinfo {author} {\bibfnamefont {D.~W.}\ \bibnamefont
  {Ferguson}}, \bibinfo {author} {\bibfnamefont {K.~N.}\ \bibnamefont {Rao}},
  \bibinfo {author} {\bibfnamefont {M.~E.}\ \bibnamefont {Mickelson}}, \ and\
  \bibinfo {author} {\bibfnamefont {L.~E.}\ \bibnamefont {Larson}},\
  }\href@noop {} {\bibfield  {journal} {\bibinfo  {journal} {J. Mol.
  Spectrosc.},\ }\textbf {\bibinfo {volume} {160}},\ \bibinfo {pages} {315}
  (\bibinfo {year} {1993})}\BibitemShut {NoStop}%
\bibitem [{\citenamefont {Dickenson}\ \emph {et~al.}(2013)\citenamefont
  {Dickenson}, \citenamefont {Niu}, \citenamefont {Salumbides}, \citenamefont
  {Komasa}, \citenamefont {Eikema}, \citenamefont {Pachucki},\ and\
  \citenamefont {Ubachs}}]{Dickenson2013}%
  \BibitemOpen
  \bibfield  {author} {\bibinfo {author} {\bibfnamefont {G.~D.}\ \bibnamefont
  {Dickenson}}, \bibinfo {author} {\bibfnamefont {M.~L.}\ \bibnamefont {Niu}},
  \bibinfo {author} {\bibfnamefont {E.~J.}\ \bibnamefont {Salumbides}},
  \bibinfo {author} {\bibfnamefont {J.}~\bibnamefont {Komasa}}, \bibinfo
  {author} {\bibfnamefont {K.~S.~E.}\ \bibnamefont {Eikema}}, \bibinfo {author}
  {\bibfnamefont {K.}~\bibnamefont {Pachucki}}, \ and\ \bibinfo {author}
  {\bibfnamefont {W.}~\bibnamefont {Ubachs}},\ }\Doi
  {10.1103/PhysRevLett.110.193601} {\bibfield  {journal} {\bibinfo  {journal}
  {Phys. Rev. Lett.},\ }\textbf {\bibinfo {volume} {110}},\ \bibinfo {pages}
  {193601} (\bibinfo {year} {2013})}\BibitemShut {NoStop}%
\bibitem [{\citenamefont {Niu}\ \emph {et~al.}(2014)\citenamefont {Niu},
  \citenamefont {Salumbides}, \citenamefont {Dickenson}, \citenamefont
  {Eikema},\ and\ \citenamefont {Ubachs}}]{Niu2014}%
  \BibitemOpen
  \bibfield  {author} {\bibinfo {author} {\bibfnamefont {M.~L.}\ \bibnamefont
  {Niu}}, \bibinfo {author} {\bibfnamefont {E.~J.}\ \bibnamefont {Salumbides}},
  \bibinfo {author} {\bibfnamefont {G.~D.}\ \bibnamefont {Dickenson}}, \bibinfo
  {author} {\bibfnamefont {K.~S.~E.}\ \bibnamefont {Eikema}}, \ and\ \bibinfo
  {author} {\bibfnamefont {W.}~\bibnamefont {Ubachs}},\ }\href@noop {}
  {\bibfield  {journal} {\bibinfo  {journal} {\jms},\ }\textbf {\bibinfo
  {volume} {300}},\ \bibinfo {pages} {44} (\bibinfo {year} {2014})}\BibitemShut
  {NoStop}%
\bibitem [{\citenamefont {Campargue}\ \emph {et~al.}(2012)\citenamefont
  {Campargue}, \citenamefont {Kassi}, \citenamefont {Pachucki},\ and\
  \citenamefont {Komasa}}]{Campargue2012}%
  \BibitemOpen
  \bibfield  {author} {\bibinfo {author} {\bibfnamefont {A.}~\bibnamefont
  {Campargue}}, \bibinfo {author} {\bibfnamefont {S.}~\bibnamefont {Kassi}},
  \bibinfo {author} {\bibfnamefont {K.}~\bibnamefont {Pachucki}}, \ and\
  \bibinfo {author} {\bibfnamefont {J.}~\bibnamefont {Komasa}},\ }\href@noop {}
  {\bibfield  {journal} {\bibinfo  {journal} {Phys. Chem. Chem. Phys.},\
  }\textbf {\bibinfo {volume} {14}},\ \bibinfo {pages} {802} (\bibinfo {year}
  {2012})}\BibitemShut {NoStop}%
\bibitem [{\citenamefont {Cheng}\ \emph {et~al.}(2012)\citenamefont {Cheng},
  \citenamefont {Sun}, \citenamefont {Pan}, \citenamefont {Wang}, \citenamefont
  {Liu}, \citenamefont {Campargue},\ and\ \citenamefont {Hu}}]{Cheng2012}%
  \BibitemOpen
  \bibfield  {author} {\bibinfo {author} {\bibfnamefont {C.-F.}\ \bibnamefont
  {Cheng}}, \bibinfo {author} {\bibfnamefont {Y.~R.}\ \bibnamefont {Sun}},
  \bibinfo {author} {\bibfnamefont {H.}~\bibnamefont {Pan}}, \bibinfo {author}
  {\bibfnamefont {J.}~\bibnamefont {Wang}}, \bibinfo {author} {\bibfnamefont
  {A.-W.}\ \bibnamefont {Liu}}, \bibinfo {author} {\bibfnamefont
  {A.}~\bibnamefont {Campargue}}, \ and\ \bibinfo {author} {\bibfnamefont
  {S.-M.}\ \bibnamefont {Hu}},\ }\href@noop {} {\bibfield  {journal} {\bibinfo
  {journal} {\pra},\ }\textbf {\bibinfo {volume} {85}},\ \bibinfo {pages}
  {024501} (\bibinfo {year} {2012})}\BibitemShut {NoStop}%
\bibitem [{\citenamefont {Tan}\ \emph {et~al.}(2014)\citenamefont {Tan},
  \citenamefont {Wang}, \citenamefont {Cheng}, \citenamefont {Zhao},
  \citenamefont {Liu},\ and\ \citenamefont {Hu}}]{Tan2014}%
  \BibitemOpen
  \bibfield  {author} {\bibinfo {author} {\bibfnamefont {Y.}~\bibnamefont
  {Tan}}, \bibinfo {author} {\bibfnamefont {J.}~\bibnamefont {Wang}}, \bibinfo
  {author} {\bibfnamefont {C.-F.}\ \bibnamefont {Cheng}}, \bibinfo {author}
  {\bibfnamefont {X.-Q.}\ \bibnamefont {Zhao}}, \bibinfo {author}
  {\bibfnamefont {A.-W.}\ \bibnamefont {Liu}}, \ and\ \bibinfo {author}
  {\bibfnamefont {S.-M.}\ \bibnamefont {Hu}},\ }\href@noop {} {\bibfield
  {journal} {\bibinfo  {journal} {\jms},\ }\textbf {\bibinfo {volume} {300}},\
  \bibinfo {pages} {60} (\bibinfo {year} {2014})}\BibitemShut {NoStop}%
\bibitem [{\citenamefont {Liu}\ \emph {et~al.}(2009)\citenamefont {Liu},
  \citenamefont {Salumbides}, \citenamefont {Hollenstein}, \citenamefont
  {Koelemeij}, \citenamefont {Eikema}, \citenamefont {Ubachs},\ and\
  \citenamefont {Merkt}}]{Liu2009}%
  \BibitemOpen
  \bibfield  {author} {\bibinfo {author} {\bibfnamefont {J.}~\bibnamefont
  {Liu}}, \bibinfo {author} {\bibfnamefont {E.~J.}\ \bibnamefont {Salumbides}},
  \bibinfo {author} {\bibfnamefont {U.}~\bibnamefont {Hollenstein}}, \bibinfo
  {author} {\bibfnamefont {J.~C.~J.}\ \bibnamefont {Koelemeij}}, \bibinfo
  {author} {\bibfnamefont {K.~S.~E.}\ \bibnamefont {Eikema}}, \bibinfo {author}
  {\bibfnamefont {W.}~\bibnamefont {Ubachs}}, \ and\ \bibinfo {author}
  {\bibfnamefont {F.}~\bibnamefont {Merkt}},\ }\href@noop {} {\bibfield
  {journal} {\bibinfo  {journal} {J. Chem. Phys.},\ }\textbf {\bibinfo {volume}
  {130}},\ \bibinfo {pages} {174306} (\bibinfo {year} {2009})}\BibitemShut
  {NoStop}%
\bibitem [{\citenamefont {Salumbides}\ \emph {et~al.}(2011)\citenamefont
  {Salumbides}, \citenamefont {Dickenson}, \citenamefont {Ivanov},\ and\
  \citenamefont {Ubachs}}]{Salumbides2011}%
  \BibitemOpen
  \bibfield  {author} {\bibinfo {author} {\bibfnamefont {E.~J.}\ \bibnamefont
  {Salumbides}}, \bibinfo {author} {\bibfnamefont {G.~D.}\ \bibnamefont
  {Dickenson}}, \bibinfo {author} {\bibfnamefont {T.~I.}\ \bibnamefont
  {Ivanov}}, \ and\ \bibinfo {author} {\bibfnamefont {W.}~\bibnamefont
  {Ubachs}},\ }\href@noop {} {\bibfield  {journal} {\bibinfo  {journal} {Phys.
  Rev. Lett.},\ }\textbf {\bibinfo {volume} {107}},\ \bibinfo {pages} {043005}
  (\bibinfo {year} {2011})}\BibitemShut {NoStop}%
\bibitem [{\citenamefont {Steadman}\ and\ \citenamefont
  {Baer}(1989)}]{Steadman1989}%
  \BibitemOpen
  \bibfield  {author} {\bibinfo {author} {\bibfnamefont {J.}~\bibnamefont
  {Steadman}}\ and\ \bibinfo {author} {\bibfnamefont {T.}~\bibnamefont
  {Baer}},\ }\Doi {http://dx.doi.org/10.1063/1.457430} {\bibfield  {journal}
  {\bibinfo  {journal} {\jcp},\ }\textbf {\bibinfo {volume} {91}},\ \bibinfo
  {pages} {6113} (\bibinfo {year} {1989})}\BibitemShut {NoStop}%
\bibitem [{\citenamefont {Niu}\ \emph {et~al.}(2015)\citenamefont {Niu},
  \citenamefont {Ramirez}, \citenamefont {Salumbides},\ and\ \citenamefont
  {Ubachs}}]{Niu2015}%
  \BibitemOpen
  \bibfield  {author} {\bibinfo {author} {\bibfnamefont {M.~L.}\ \bibnamefont
  {Niu}}, \bibinfo {author} {\bibfnamefont {F.}~\bibnamefont {Ramirez}},
  \bibinfo {author} {\bibfnamefont {E.~J.}\ \bibnamefont {Salumbides}}, \ and\
  \bibinfo {author} {\bibfnamefont {W.}~\bibnamefont {Ubachs}},\ }\Doi
  {http://dx.doi.org/10.1063/1.4906244} {\bibfield  {journal} {\bibinfo
  {journal} {\jcp},\ }\textbf {\bibinfo {volume} {142}},\ \bibinfo {eid}
  {044302} (\bibinfo {year} {2015})}\BibitemShut {NoStop}%
\bibitem [{\citenamefont {Xu}\ \emph {et~al.}(2000)\citenamefont {Xu},
  \citenamefont {van Dierendonck}, \citenamefont {Hogervorst},\ and\
  \citenamefont {Ubachs}}]{Xu2000}%
  \BibitemOpen
  \bibfield  {author} {\bibinfo {author} {\bibfnamefont {S.}~\bibnamefont
  {Xu}}, \bibinfo {author} {\bibfnamefont {R.}~\bibnamefont {van Dierendonck}},
  \bibinfo {author} {\bibfnamefont {W.}~\bibnamefont {Hogervorst}}, \ and\
  \bibinfo {author} {\bibfnamefont {W.}~\bibnamefont {Ubachs}},\ }\href@noop {}
  {\bibfield  {journal} {\bibinfo  {journal} {\jms},\ }\textbf {\bibinfo
  {volume} {201}},\ \bibinfo {pages} {256} (\bibinfo {year}
  {2000})}\BibitemShut {NoStop}%
\bibitem [{\citenamefont {Bailly}\ \emph {et~al.}(2010)\citenamefont {Bailly},
  \citenamefont {Salumbides}, \citenamefont {Vervloet},\ and\ \citenamefont
  {Ubachs}}]{Bailly2009}%
  \BibitemOpen
  \bibfield  {author} {\bibinfo {author} {\bibfnamefont {D.}~\bibnamefont
  {Bailly}}, \bibinfo {author} {\bibfnamefont {E.}~\bibnamefont {Salumbides}},
  \bibinfo {author} {\bibfnamefont {M.}~\bibnamefont {Vervloet}}, \ and\
  \bibinfo {author} {\bibfnamefont {W.}~\bibnamefont {Ubachs}},\ }\href@noop {}
  {\bibfield  {journal} {\bibinfo  {journal} {\molp},\ }\textbf {\bibinfo
  {volume} {108}},\ \bibinfo {pages} {827} (\bibinfo {year}
  {2010})}\BibitemShut {NoStop}%
\bibitem [{\citenamefont {Pachucki}\ and\ \citenamefont
  {Komasa}(2015)}]{Pachucki2015}%
  \BibitemOpen
  \bibfield  {author} {\bibinfo {author} {\bibfnamefont {K.}~\bibnamefont
  {Pachucki}}\ and\ \bibinfo {author} {\bibfnamefont {J.}~\bibnamefont
  {Komasa}},\ }\href@noop {} {\bibfield  {journal} {\bibinfo  {journal}
  {\jcp},\ }\textbf {\bibinfo {volume} {143}},\ \bibinfo {eid} {034111}
  (\bibinfo {year} {2015})}\BibitemShut {NoStop}%
\end{thebibliography}%

\end{document}